\documentstyle[prl,tighten,aps,multicol,amstex,prabib,epsfig]{revtex}
\newcommand{\1}{|1\rangle}
\newcommand{\0}{|0\rangle}

\newcommand{\dagg}[1]{#1^{\dag}}
\newcommand{\opn}[1]{\operatorname{#1}}

\newenvironment{thm}{\noindent \textbf{Proposition:} \rm}{\\}
\newenvironment{pf}{\textit{Proof: } \rm}{\mbox{}\hfill $\Box$\\}
\newenvironment{lemma}{\noindent \textbf{Lemma:} \rm}{\\}

\begin{document}
\title{Entanglement Optimization for Pairs of Qubits}
\author{Juliane Strassner and Christopher Witte}
\address{Institut f{\"u}r Theoretische Physik,Technische Universit{\"a}t Berlin, D-10623 Berlin,
Germany}

\date{\today}
\maketitle
\begin{abstract}
Local Operations enhancing the entanglement  of bipartite quantum
states are of great interest in quantum information processing.
Subject of this paper are local selective operations acting on
single copies of states. Such operations can lead to larger
entanglement with respect to a certain measure as studies before by
the Horodeckis and A.~Kent et al. \cite{HHH98b,K98,KLM98}. We
present a complete characterisation of all local operations yielding
optimal entanglement for pairs of qubits, extending
former results of A.~Kent et al. We introduce a new
technique for the classification of states according to their
behaviour under entanglement optimizing operations, using the
entanglement properties of the support of density matrices.
\end{abstract}
\begin{multicols}{2}
The key resource of quantum information theory is quantum
entanglement. It is of vital importance to find a measure for
entanglement and investigate how it can be increased for a given
state. Many applications of quantum information theory, like
teleportation, dense coding or quantum cryptography, require
maximally entangled states of two qubits shared by two distinct
parties, traditionally called Alice and Bob.
Maximally entangled Pairs, however, have to be prepared in a common quantum process at a certain place.
In order to share
them between Alice and Bob at least one of the two particles must be sent
through a quantum channel (e.g.~an optical fibre). During this
process interaction with the environment leads to a loss of
entanglement, the state evolves to a non-maximally entangled mixed
state.

Therefore purification, also called distillation, becomes necessary:
a process which increases the entanglement of given pairs by local
operations and classical communication (LQCC) performed by Alice and
Bob. The first purification protocol was presented by Bennett et
al.\cite{BBPS96}. It involves local operations acting on many pairs,
called collective operations. Using such a collective scheme one can
obtain pure maximally entangled particles from any given
two-spin-$\frac{1}{2}$-state, even from mixed ones.

Recent publications raised the question, if it is also possible to
purify arbitrary two-spin-$\frac{1}{2}$-states if only non-collective
operations are allowed, i.e.~operations that act on each pair
individually. It was shown that this is possible only for pure states.
Mixed states of two qubits cannot be purified to maximally entangled
states by local operations. Nevertheless there are mixed states
whose entanglement can be increased by such protocols. For some of
these states it is even possible to reach maximal entanglement, but only  with
vanishing probability, i.e.~there is no limit for the entanglement
that can be distilled from these states. This process was introduced
by Horodecki et al. \cite{HHH98b} and called {\it
quasi-distillability}.

In this letter we classify states of two qubits with respect to their
maximal distillable entanglement. First we present the entanglement
properties of the support of the considered density matrices, which
will lead to the classification given below. The
proofs will be given in the last part of this letter.

We use the following facts (see \cite{Woo98,K98,KLM98}):

\begin{itemize}
   \item The considered distillation protocols involve LQCC's of a
     special form: after the operation the state $\rho$ takes the form
     \begin{eqnarray}
     \label{eq:1}
     \Theta(\rho)=\frac{A\otimes B \rho A\otimes
       B}{\textrm{Tr}(A\otimes B \rho A\otimes B)}
     \end{eqnarray}
   \item The entanglement of formation ($E_{F}$) of a state of two
     qubits can be calculated by the formula of Wootters as
     \[E_{F}(\rho)=H(\frac{1+\sqrt{1-C^{2}(\rho)}}{{2}})\] with $H(p)=p
     \log_{2}p-(1-p)\log_{2}(1-p)$\\ and  the concurrence $C$, which is
     given by
     $C=\max\{0,\lambda_{1}-\lambda_{2}-\lambda_{3}-\lambda_{4}\}$,
     the difference of the eigenvalues $\lambda_{i}$ of $\tilde \rho
     \rho=\sigma_{2}\otimes\sigma_{2}\bar{\rho}\sigma_{2}\otimes\sigma_{2}\rho$,
     where the eigenvalues are taken in decreasing order.
   \item Linden, Massar and Popescu have shown in \cite{LMP98} that
     under operations of the form (\ref{eq:1}) the eigenvalues $\lambda_{i}$ of the
     matrix $\tilde \rho \rho$ transform as
      \begin{eqnarray}
      \label{eq:2}
      \lambda_{i}^{'}=
c_{1}\lambda_{i}
      \end{eqnarray}
     with a factor $c_{1}$ that is independent of $i$.

As a consequence the ratio of the eigenvalues is constant under local
operations.
\end{itemize}
We will show that
the main qualitative feature of density matrices with
respect to entanglement optimization procedures is the number of
product vectors in the \emph{support} of this matrix.

The \emph{support} of a density matrix is the orthogonal complement
of its \emph{kernel}, i.e. for selfadjoint matrices identical with
its \emph{range}. We can also think of the support as the linear
span of the eigenvectors with non-vanishing eigenvalues. This subspace of ${\mathbb{C}}^{2}
\otimes  {\mathbb{C}}^2$ determines a subset of the state space
containing all states having support included in this subspace. Such
a subset of state space is called a \emph{face} in convex analysis.
The different faces of a state space over a
tensor product Hilbert space can be characterised firstly by the
dimension of the respective subspace (i.e. the rank of the generic
elements) and secondly by the structure of the subset of product
vectors in this subspace.  The first property is invariant under all
linear isomorphisms of the Hilbert space. The second one is
invariant under \emph{factorising} linear isomorphisms.

In the simplest case of two qubits a complete classification can
be done. For that purpose we have a look at all possible subspaces
$U\subseteq{\mathbb{C}}^{2} \otimes  {\mathbb{C}}^2$ (see also \cite{Lew} for some details):
\renewcommand{\labelenumi}{\Alph{enumi}.}
\begin{enumerate}
    \item \label{sup1}The case $\dim U=4$ is trivial. There is only one such
    subspace: ${\mathbb{C}}^{2} \otimes  {\mathbb{C}}^2$ itself.
    \item \label{sup2}For $\dim U=3$ there are two cases, easily characterised by
    their one dimensional orthogonal complement (the \emph{kernel} of
    the respective matrices). Either the complement is factorising and
    the subspace contains two factorising hyperplanes, or the
    complement is entangled and the subspace contains a cone-shaped
    set of factorising vectors.
            \epsfig{file=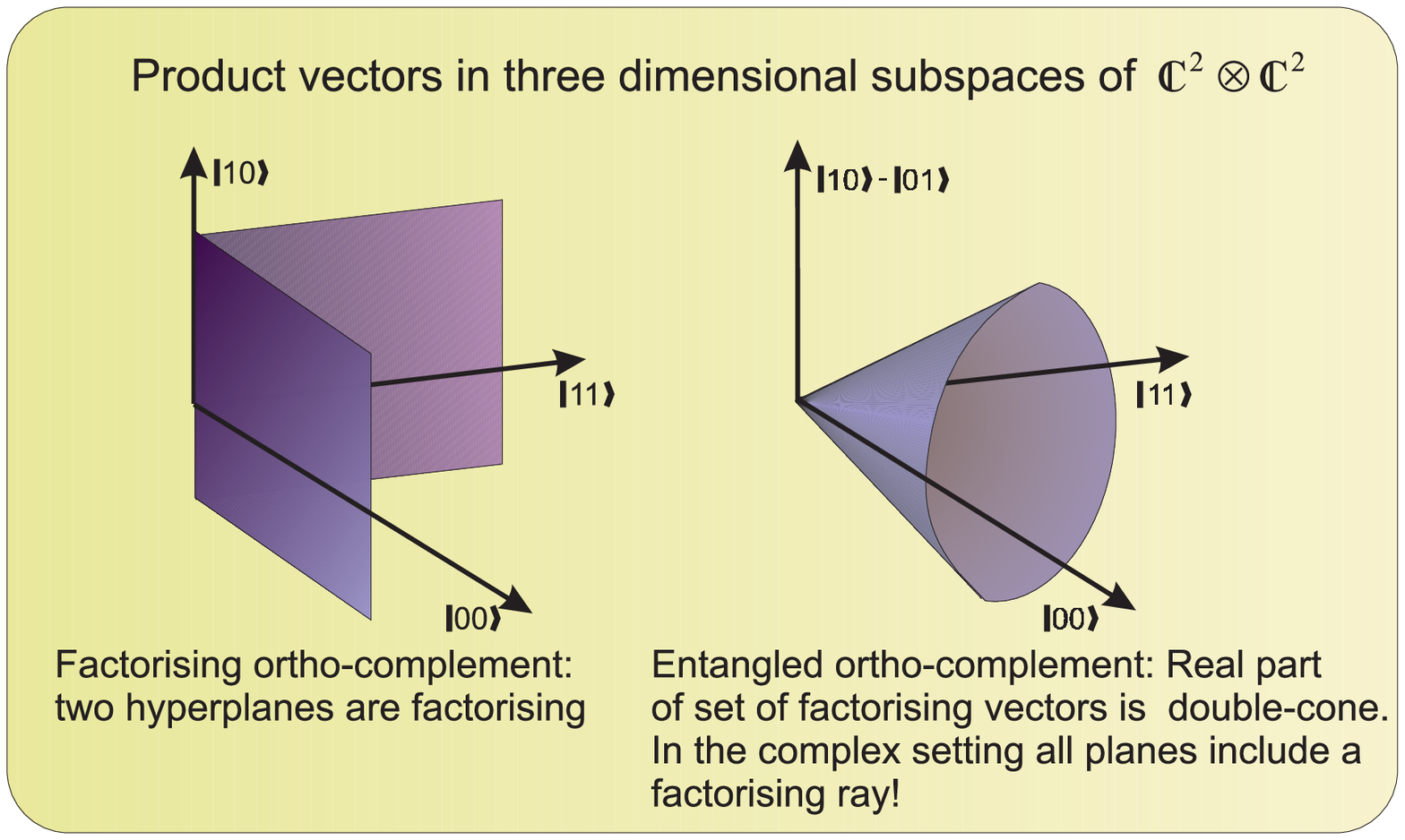,height=42mm}
    \item \label{sup3}$\dim U=2$ offers three possibilities: a) the whole subspace is
    factorising, b) exactly two linear independent vectors are
    factorising and c) exactly one linear independent vector
    factorises.
            \epsfig{file=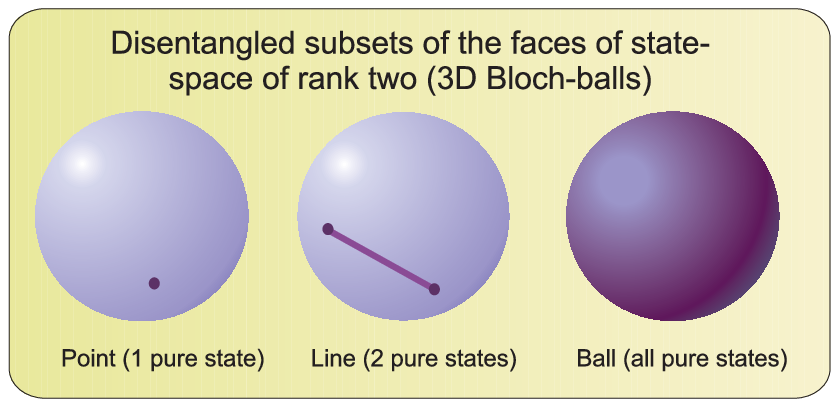,height=34mm}
    \item \label{sup4}$\dim U=1$ is trivial again. Either it factorises, or not.
\end{enumerate}
These properties of the support of density matrices will now be
shown to be the only features determining the behaviour under
local transformations with respect to entanglement optimization.\\
We state our results as follows:

Let $\rho$ be an entangled state on ${\mathbb{C}}^{2} \otimes
{\mathbb{C}}^2$.
 Set $$M_\rho:=\sup \{E_F(\Theta(\rho)): \Theta \mbox{ local operation }\}. $$
If $\rho$ is Bell-diagonal, it has been shown in \cite{K98} that no
local operation can increase the Entanglement of Formation $E_F(\rho)$ of the state, i.e.
$M_\rho=E_F(\rho)$. Thus we assume here that $\rho$ be \textbf{not}
Bell-diagonal. Then one of the following holds true:
\renewcommand{\labelenumi}{\Roman{enumi}.}
 \begin{enumerate}
        \item $\opn{rank} \rho=4$. As seen in \cite{K98} there
        exists a local operation $\Xi$, such that
        $E_F(\Xi(\rho))=M_\rho<1$ and $\Xi(\rho)$ is Bell-diagonal. We call
        this property  of the state
        \emph{incomplete distillability}.
        \item $\opn{rank} \rho=3$. $M_\rho<1$. Either (i.) the kernel of $\rho$
        contains no product vector and a local operation $\Xi$ exists, such that
        $E_F(\Xi(\rho))=M_\rho$ and $\Xi(\rho)$ is Bell-diagonal, i.e.
        $\rho$ is \emph{incompletely distillable}.
        Or (ii.) the kernel of $\rho$ is the linear span of a product vector
        and there is no local operation with $E_F(\Xi(\rho))=M_\rho$.
        In this case we can find a sequence of local operations $\{\Xi_n\}_n$,
        with $\sigma=\lim_{n\to \infty} \Xi_n(\rho)$  Bell-diagonal and
        $E_F(\sigma)=M_\rho$. These states will be called
        \emph{incompletely quasi-distillable}.

        \item $\opn{rank} \rho=2$. Either (i.) the support of $\rho$
        contains exactly one linear independent product vector. In this case $M_\rho=1$,
        but there is no local operation with $E_F(\Xi(\rho))=1$. Instead we find
         a sequence of local operations $\{\Xi_n\}_{n\in\mathbb{N}}$,
        with $\lim_{n\to \infty} \Xi_n(\rho)$  a Bell-state
        (\emph{i.e. }$\rho$ is \emph{ quasi-distillable}).
        Or (ii.) the support
        of $\rho$ contains exactly two linear independent product vectors
        and it exists  a local operation $\Xi$, such that
        $E_F(\Xi(\rho))=M_\rho$ and $\Xi(\rho)$ is Bell-diagonal, i.e. the state
        is \emph{incompletely distillable}.
        \item $\opn{rank} \rho=1$, i.e. the state is pure: $M_\rho=1$
          Then $\rho$
        ist \emph{distillable}, i.e.
        there is a local operation $\Xi$, such that $\Xi(\rho)$
        is a Bell-state.
  \end{enumerate}
In order to prove our programme we have a look at the previously
existing results. Kent et al. have shown in \cite{KLM98} that states
can be brought to a Bell-diagonal form with optimal entanglement, as
long as the quantity $\opn{Tr}(A \otimes B \rho
\dagg{A}\otimes\dagg{B})$ cannot be zero for those states.
Nevertheless for states that do not fulfil this assumption, like
quasi-distillable states, an essential continuity argument fails. For
that reason we will now characterize quasi-distillable and
incompletely quasi-distillable states, which will lead to the
characterization given above.\\
\begin{lemma}
A quasi-distillable  or incompletely quasi-distillable state has a
factorising vector in its kernel.
\end{lemma}

\begin{pf}
  If $\rho$ is quasi-distillable or incompletely quasi-distillable,
  then the probability for reaching the product $\Xi_{n}(\rho)$ tends to
  $0$ as $n\to \infty$.

In the following we will use the operators
  $\tilde A_{n} \otimes \tilde B_{n}:=\frac{ A_{n}\otimes
  B_{n}}{\| A_{n}\otimes  B_{n}\|}$, which produce the same
state $\Xi_{n}(\rho)$. The probability of getting the state
$\Xi_{n}(\rho)$ is given by
\[p_{n}:=\textrm{Tr}(\tilde A_{n} \otimes \tilde B_{n}\rho \tilde
A_{n}^{\dagger}\otimes \tilde B_{n}^{\dagger}).\] Since $\rho$  is
quasi-distillable, it follows, that
\[  \lim_{n\to\infty}\textrm{Tr}(\tilde A_{n} \otimes \tilde B_{n}\rho
\tilde A_{n}^{\dagger}\otimes \tilde
B_{n}^{\dagger})=\lim_{n\to\infty}\textrm{Tr}(\rho \tilde
A^{\dagger}_{n} \tilde A_{n} \otimes \tilde B^{\dagger}_{n} \tilde
B_{n})=0.\] Since $\|\tilde A_{n} \otimes \tilde B_{n}\|=1$
the set of these operators is compact and there exists a  subsequence
$\tilde A_{n_{i}} \otimes \tilde B_{n_{i}}$ that tends to a limit
operation $\tilde A \otimes \tilde B:=\lim_{i\to\infty}\tilde
A_{n_{i}} \otimes \tilde B_{n_{i}}$ such that $\|\tilde A
\otimes \tilde B\|=1.$ Then it follows that
\begin{eqnarray}
  \label{eq:5}
&\lim_{i\to\infty}&\textrm{Tr}(\tilde A_{n_i} \otimes \tilde B_{n_i}\rho
\tilde A_{n_i}^{\dagger}\otimes \tilde B_{n_i}^{\dagger})\nonumber\\ &=&
\textrm{Tr}(\tilde A \otimes \tilde B\rho \tilde A^{\dagger}\otimes
\tilde B^{\dagger})\nonumber\\ &=&\textrm{Tr}(\rho \tilde
A^{\dagger} \tilde A \otimes \tilde B^{\dagger} \tilde B)=0\nonumber
\end{eqnarray}
The operators $\tilde A^{\dagger}\tilde A,\tilde B^{\dagger}\tilde B$ are positive and can in
order of the spectral theorem be decomposed by
\[ \tilde A^{\dagger} \tilde A=\sum_{i}a_{i}P_{i}, \hspace{1cm}\tilde B^{\dagger} \tilde
  B= \sum_{j}b_{j}Q_{j}\]
with  $a_{i},b_{j}\in{\mathbb{R}}^{+}$ and projectors
$P_{i}=|\phi_{i}\rangle\langle\phi_{i}|,Q_{j}=|\psi_{j}\rangle\langle\psi_{j}|$.
The above equation then takes the form
\begin{eqnarray}
  \label{eq:75}
  &&\textrm{Tr}(\rho \tilde A^{\dagger} \tilde
    A\otimes\tilde B^{\dagger} \tilde B)=0\nonumber\\
  &\Leftrightarrow&\sum_{i,j}\textrm{Tr}(\rho a_{i}b_{j}P_{i}\otimes Q_{j})=0\nonumber\\
  &\Leftrightarrow&\sum_{i,j}a_{i}b_{j}\langle\phi_{i}\otimes\psi_{j}|\rho|\phi_{i}\otimes\psi_{j}\rangle=0\nonumber\\
&\Rightarrow&\langle\phi_{i}\otimes\psi_{j}|\rho|\phi_{i}\otimes\psi_{j}\rangle=0\hspace{0.5cm}\textrm{for
  all}\hspace{0.5cm}a_{i},b_{j},\hspace{0.5cm}
a_{i}b_{j}\neq 0\nonumber\\ &\Rightarrow&
\rho^{\frac{1}{2}}|\phi_{i}\otimes\psi_{j}\rangle=0\hspace{1.6cm}\textrm{for
  all}\hspace{0.5cm}a_{i},b_{j},\hspace{0.5cm}
a_{i}b_{j}\neq 0.\nonumber
\end{eqnarray}

The $|\phi_{i}\otimes\psi_{j}\rangle$ are then part of the kernel of
$\rho^{\frac{1}{2}}$ for all $a_{i},b_{j}$ with $a_{i}b_{j}\neq 0$
and also part of the kernel of $\rho$. Since the operation $\tilde
A\otimes \tilde B$ cannot be zero there has to be a product
$a_{i}b_{j}\neq 0$ and therefore the corresponding product vector
$|\phi_{i}\otimes\psi_{j}\rangle$ lies in the kernel of $\rho$.
\end{pf}

\begin{thm}
A density matrix $\rho$ on ${\mathbb{C}}^2 \otimes {\mathbb{C}}^2$
is quasi-distillable if and only if it has rank two and the
support of $\rho$ is spanned by a factorising vector $\phi_1
\otimes \phi_2$ and an orthogonal entangled vector $\psi$.
\end{thm}

\begin{pf}
  For any quasi-distillable state $\rho$ there exists a sequence of operations
   $\{\Xi_n\}_{n\in {\mathbb{N}}}$ with
   $\lim_{n\to\infty}\Xi_n(\rho)=\rho_\infty$ where $\rho_\infty$ is a
   Bell-state. Since the vector
   $\vec{\lambda}_{\infty}$ containing the
   eigenvalues  of $\tilde\rho_\infty \rho_\infty$  is given
   by $(1,0,0,0)$ and by (\ref{eq:2}) the ratio of eigenvalues cannot change under the
   local operation $\Xi_n$, the vector $\vec{\lambda}_{0}$ belonging to the original
   state $\rho$ must be of the
   form $(\lambda_{0},0,0,0)$, with $\lambda_{0}\neq 0$,
   i.e.~the rank of $\tilde\rho\rho$ is $1$.

   Since $\rho$ is quasi-distillable, there is a  factorising
   vector $\phi_3\otimes\phi_4$ in its kernel. In the tensor basis $(|00\rangle,|01\rangle,|10\rangle,|11\rangle)$ we can assume without loss
of generality that $\phi_3\otimes\phi_4$ have the form $|11\rangle$. The
matrix form of $\rho$ is in that case
\begin{equation}\label{eq:7} \rho=\begin{pmatrix}
 1-a-d  & e& c & 0 \\
  \bar{e} &a & b & 0\\
  \bar{c}& \bar{b}& d &0 \\
    0&0&0&0
\end{pmatrix}.\end{equation}
Since $\rho$ ist positive,
we have $a,d\geq 0$ and
the eigenvalues of the matrix $\tilde{\rho}\rho $ are
$(0,0,(\sqrt{ad} -|b|)^{2},(\sqrt{ad} +|b|)^{2})$.
Since $\tilde{\rho}\rho $ is of rank 1, we get $\sqrt{ad}=|b|$.
This leads to
\[\rho=\left(  \begin{array}{cccc}
1-a-d&e^{i\delta}c\sqrt{\frac{a}{d}}&c&0\\
e^{-i\delta}\bar{c}\sqrt{\frac{a}{d}}&a&e^{i\delta}\sqrt{ad}&0\\
\bar{c}&e^{-i\delta}\sqrt{ad}&d&0\\ 0&0&0&0\\
  \end{array}\right)\]
where the condition $ (1+a+d)d\leq
|c|^{2}\leq(1+a+d)d+\frac{d}{4(a+d)}$ has to be
made for $\rho$ to be positive.

The support of $\rho$ is therefore the subspace spanned by the
vectors $|00\rangle$ and
$\frac{1}{\sqrt{a+d}}(\sqrt{a}|01\rangle+\sqrt{d}|10\rangle)$.

These states are indeed quasi-distillable, which can be shown by
using the following protocol:
\begin{enumerate}
\item First Alice applies the local filtration
  $A=\textrm{diag}(\sqrt{d},\sqrt{a})$.
\item  Now Alice and Bob apply the local operation given by
\begin{eqnarray}
  \label{eq:70}
  A_{n}=\left(
  \begin{array}{cc}
\frac{1}{n} & 0\\ 0 & 1
  \end{array}\right),&&
B_{n}=\left(
  \begin{array}{cc}
\frac{1}{n} & 0\\ 0 & 1
  \end{array}\right).
\end{eqnarray}
\end{enumerate}

This operation produces the state
\begin{eqnarray}
  \label{eq:6}
\rho_{n}&=&\frac{A_{n}\otimes B_{n}\rho^{'} A_{n}\otimes
  B_{n}}{\textrm{Tr}(A_{n}\otimes B_{n}\rho^{'} A_{n}\otimes B_{n})}\nonumber\\
&=&c_{1} \left(\begin{array}{cccc} \frac{1-a-d}{1-a-d+2a
  n^{2})}&\frac{e^{i\delta}c}{(\frac{1-a-d}{n}+2a n)}&\frac{c\sqrt{ad}}{(\frac{1-a-d}{n}+2a n)d}&0\\
\frac{e^{-i\delta}\bar{c}}{(\frac{1-a-d}{n}+2a
n)}&\frac{a}{\frac{1-a-d}{n^{2}}+2a}&\frac{e^{i\delta}a}{\frac{1-a-d}{n^{2}}+2a}&0\\
\frac{\bar{c}\sqrt{ad}}{(\frac{1-a-d}{n}+2a
n)d}&\frac{e^{-i\delta}a}{\frac{1-a-d}{n^{2}}+2a}&\frac{a}{\frac{1-a-d}{n^{2}}+2a}&0\\
0&0&0&0\\
  \end{array}\right)\nonumber
\end{eqnarray}
with $c_{1}=\frac{1}{d(1+a-d)}$.

 It can be shown that \[\lim_{n\to\infty}\rho_{n}=
\rho_{\infty}=|01\rangle+e^{i\lambda}|10\rangle.\] This is a
maximally entangled state and thus $\rho$ is quasi-distillable.
\end{pf}

\begin{thm}
A density matrix $\rho$ on ${\mathbb{C}}^2 \otimes {\mathbb{C}}^2$ is
incompletely quasi-distillable if and only if the kernel of $\rho$
is spanned by a factorising vector $\phi_1\otimes\phi_2$. The best
distillation product of $\rho$ is in that case a Bell diagonal state
of rank two.
\end{thm}

\begin{pf}
Again we assume without loss of generality, that the state can be
written as
 (\ref{eq:7}).
 We know from
\cite{K98} that the entanglement of formation of a state can be
further increased if and only if it is not Bell diagonal. An optimal
distillation product must be thus Bell diagonal. From this and the
fact that the distillation product must still have a factorising
vector in its kernel, it follows that such a \textit{final} state
must be of the form
\[ \sigma=\begin{pmatrix}
  0 & 0& 0 & 0 \\
  0 &1/2 & f & 0\\
  0& \bar{f}& 1/2 &0 \\
    0&0&0&0
\end{pmatrix},\]
which yields for $\sigma \tilde{\sigma}$ eigenvalues
$(0,0,(1/2+|f|)^2,(1/2-|f|)^2)$. Since by \cite{KLM98}the ratio of
these eigenvalues cannot change under local operations, we see that
\[\frac{\sqrt{a d}+|b|}{\sqrt{a d}-|b|}=\frac{1/2+|f|}{1/2-|f|}\]
must be fulfilled. From this we get
\[\frac{|b|}{2\sqrt{ad}}=|f|.\]
The rank of $\sigma$ is two whereas $\rho$ has rank three, because
it is neither distillable nor quasi-distillable. For that reason no
local operation can transform $\rho$ into the entangled state
$\sigma$, since such a local operation would have to decrease the
rank (thus being not one to one) and have entangled vectors in its
range (see \cite{K98}). Nevertheless there is a sequence of
operations which with probability tending to zero yields $\sigma$ in
the limit. For this purpose we use the operation elements
\[A_n:=\begin{pmatrix}
  \frac{1}{n} & 0 \\
  0 &\frac{1}{\sqrt{d}}
\end{pmatrix}\quad B_n:=\begin{pmatrix}

  \frac{1}{n} & 0 \\
  0 &\frac{1}{\sqrt{a}}
\end{pmatrix}\] getting
\begin{eqnarray}
\rho_n&:=&A_n\otimes B_n \rho A_n\otimes B_n \\
&=&\begin{pmatrix}
 \frac{1-a-d}{n^4}  & \frac{e}{n^3\sqrt{a}} & \frac{c}{n^3\sqrt{d} } & 0 \\
  \frac{\bar{e}}{n^3\sqrt{a}} &\frac{1}{n^2} & \frac{b}{n^2\sqrt{ad}} & 0\\
  \frac{\bar{c}}{n^3\sqrt{d} }& \frac{\bar{b}}{n^2\sqrt{ad}}& \frac{1}{n^2} &0 \\
    0&0&0&0
\end{pmatrix}.
\end{eqnarray}

We easily see that the limit of the normalized operation (only
leaving those terms proportional to $1/n^2$) is the state $\sigma$.
\end{pf}

\begin{thm}
An entangled density matrix $\rho$ on ${\mathbb{C}}^2 \otimes
{\mathbb{C}}^2$ of rank two is incompletely distillable if and
only if the support of $\rho$ is spanned by two factorising
vectors $\phi_1 \otimes \phi_2$ and  $\psi_1 \otimes \psi_2$.
\end{thm}

\begin{pf}
Since $\rho$ is entangled, there must be an entangled vector in its
support. If there would be only one such vector, $\rho$ would be quasi-distillable (see above).
For that reason there must be {\it exactly} two linear
independent factorising vectors in its support, i.e. no linear
combination  of these two factorises . Choosing without loss of
generality $\phi_1 \otimes \phi_2=\0 \1$ we see that $\psi_1 \otimes
\psi_2=(\alpha \0+ \beta \1)  \otimes (\gamma \0 + \delta \1)$, where
$\beta \neq 0$ and $\gamma \neq 0$. The invertible local operation
$$\Psi \mapsto \begin{pmatrix} 1 & \frac{-\alpha}{\beta} \\ 0 & \frac{1}{\beta}
\end{pmatrix} \otimes
\begin{pmatrix} \frac{1}{\gamma} & 0 \\ \frac{-\delta}{\gamma} & 1\end{pmatrix} \Psi $$
leaves $|01\rangle$ invariant, but transforms $\psi_1 \otimes \psi_2$ into
$|10\rangle$. This means we can assume $\psi_1 \otimes \psi_2=|10\rangle$
without changing the properties of the state with respect to local
transformations {\it qualitatively}.   In this case, following the
notation of the preceding proof, the density matrix reads:
\[ \rho=\begin{pmatrix}
 0  & 0 & 0 & 0 \\
  0 & a & b & 0\\
  0& \bar{b}& d &0 \\
    0&0&0&0
\end{pmatrix}.\]
The final argument follows  the preceding proof, yielding a
distillation product $\sigma$ as seen above. Nevertheless there is a
minor difference: since $\rho$ has rank two, we can now find  a {\it
direct} transformation $\rho \mapsto \sigma$, which clearly must be
given by the operation elements:
\[A \otimes B:=\frac{1}{\sqrt{2}}\begin{pmatrix}
  1 & 0 \\
  0 &\frac{1}{\sqrt{d}}
\end{pmatrix}\otimes \begin{pmatrix}

 0 & 0 \\
  0 &\frac{1}{\sqrt{a}}
\end{pmatrix}.\]
\end{pf}

We have seen that the appropriate method to study single-copy
distillation procedures is looking at the entanglement properties of
the state's support. These properties decide whether a state can be
brought directly into Bell-diagonal form having optimal entanglement
of formation or whether this can be achieved only in a limiting
process with vanishing probability in the limit. It is an obvious
task to generalize these ideas to higher dimension. It must be
mentioned, nevertheless, that concepts like {\it unextendible product
bases} will play a major role and will make such an analysis far
more sophisticated.

\end{multicols}
\end{document}